\documentclass{osa-article}

\journal{osajournal}

\articletype{Research Article}

\begin{document}

\title{A high sensitivity tool for geophysical applications: A geometrically locked Ring Laser Gyroscope.}

\author{
E.~Maccioni,\authormark{1,2} N.~Beverini,\authormark{1} G.~Carelli,\authormark{1,2,*} G.~Di~Somma,\authormark{1,2} A.~Di~Virgilio,\authormark{2} and P.~Marsili\authormark{1,2}
}

\address{
\authormark{1}University of Pisa, Pisa, Italy,

\authormark{2}INFN Sezione di Pisa, Pisa, Italy
}

\email{\authormark{*}giorgio.carelli@unipi.it}

\begin{abstract}
 This work demonstrates that a middle size ring laser gyroscope (RLG)
 can be a very sensitive and robust instrument for rotational seismology, 
 even if it operates in a quite noisy environment. The RLG has a 
 square cavity, $1.60\times 1.60$ m$^2$, and it lies in a plane 
 orthogonal to the Earth rotational axis. The Fabry-Perot optical 
 cavities along the diagonals of the square were accessed and their 
 lengths were locked to a reference laser. Through a  quite simple  
 locking circuit, we were able to keep  the sensor fully operative for 
 14~days. We verified that the prototype property 
 are compatible with the seismic requirements.
The obtained long term stability is of the order of 3~nanorad/s and the 
 short term sensitivity close is to 2~nanorad/s$\cdot$Hz$^{-1/2}$ . 
 These results are limited only by the noisy environment, 
 our laboratory is located in a building downtown.
\end{abstract}

\section{Introduction}
	
Sensing rotational motions has a wide range of applications, from navigation tasks in flight and space operations, to measuring Earth’s rotation rate, rotational ground motions due to earthquakes, and vibrations of buildings\cite{passaro2017}.
Gyroscopes are devices mounted on a frame and able to sense an angular velocity if the frame
is rotating. Many classes of gyroscopes exist with very different performance,
depending on the operating physical principle and the involved technology. 
The most popular are  mechanical gyroscopes \cite{britting1971}, optical gyroscopes,
both active and passive\cite{SCHREIBERCR, schreiber2009b,  bernauer2018, Liu2019, zou21},
and Micro-electromechanical system (MEMS) gyroscopes\cite{greiff1991, jia2021}.
Optical gyroscopes are based on Sagnac effect. Passive fibre-optic (FOG) gyros 
or active ring laser gyros were built, both classes outperform mechanical devices by orders of magnitude and are the technical choice for high-resolution, broadband observations of rotational motions in geodesy and geophysics and seismology \cite{SCHREIBERCR, Schreiber2006}. 
FOG gyroscopes can be used for structural engineering or seismic applications
and they can reach sensor self-noise lower than $30$~nrad$^{-1}$~Hz$^{-1/2}$,
but since a FOG measures a phase shift between two beams, 
its accuracy is limited\cite{bernauer2018}.
The output of an active ring laser is the beat frequency of two counter-propagating
laser beams  that is directly proportional to the rotation rate 
around an axis orthogonal to the plane defined by the laser beams\cite{SCHREIBERCR},
the reached sensitivity is of the order of picorad/s/Hz \cite{Schreiber2011, zou21}

Large RLG (with a perimeter of several meters) are capable to measure angular rotations with precision better that a fraction of prad/s, not far from what is necessary for General Relativity tests 
(about $10^{-14}$~rad/s)\cite{angela2017}.
They are able to combine the status of art 
sensitivity with a bandwidth above 100~Hz, long term operation
and very large dynamic range.
The same device can efficiently records extremely low amplitude 
signals and large shocks.
RLG are based on the Sagnac effect that appears as a difference in 
the optical path between waves propagating in opposite direction in a 
rotating closed loop. As a consequence, in a rotating RLG  a difference  
arises between the frequencies emitted by the laser in the two opposite 
directions (Sagnac frequency). For a RLG rigidly connected to the ground,
the Earth rotation velocity is by far the dominant component of~$\Omega$.

  Eq. (\ref{eq:general}) gives the general relation connecting 
  the Sagnac frequency $f_s $,
  and the modulus of the local angular rotation rate $\Omega$\cite{RSIUlli}:
  \begin{equation}
      f_s = 4 \frac{A}{P \lambda} \Omega\cdot \cos(\theta)
      \label{eq:general}
  \end{equation}
  
where $A$ is the area enclosed by the optical path, 
$P$ is the ring perimeter length, 
$\lambda$ the intracavity laser wavelength, and
$\theta$ is the angle between the area vector and the rotational axis.
   
An extremely sensitive ring laser system (G-ring) was installed in 2002 at the Geodetic Observatory of Wettzell\cite{Schreiber2009} measuring the local component of rotation around the vertical axis. The
G-ring was specifically designed for geodesy, built on a monolithic Zerodur structure, buried underground, thus providing sufficient long-term stability to be able to resolve tidal effects and polar motion, (e.g. \cite{schreiber2003, Schreiber2011}.
ROMY, a multi-component ring laser system with higher sensitivity for each component, was recently developed\cite{igel2021} for geodesy.

\begin{figure}[htbp]
\begin{center}
\includegraphics[width=\columnwidth]{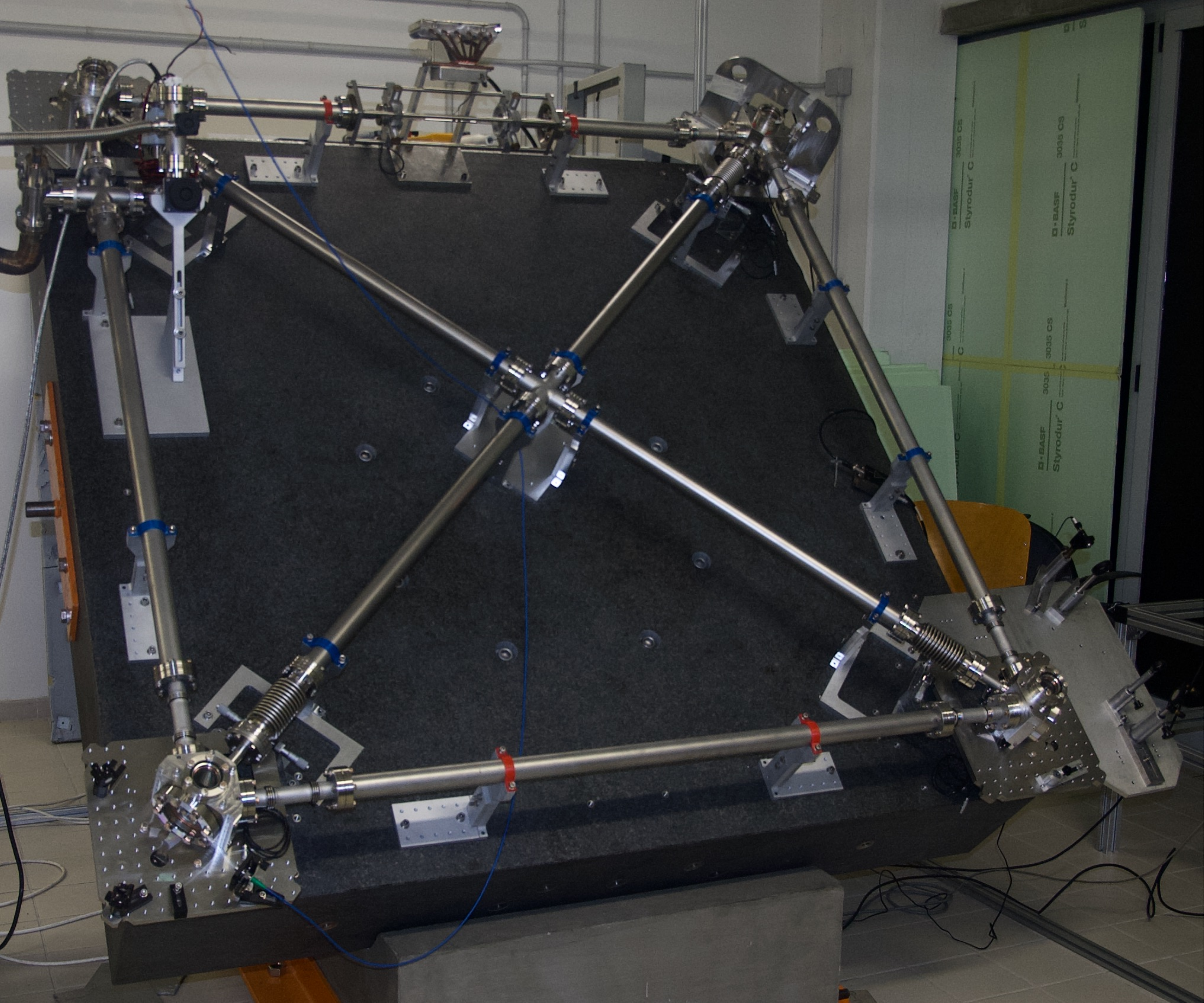}
\caption{GP2 gyroscope inside the building of the   INFN Pisa section}
\label{photo}
\end{center}
\end{figure}

Our prototype, GP2, is a square laser gyroscope 
with a perimeter of 6.40~m, that is operative 
in a laboratory placed inside the basement of the Pisa INFN building, 
see Fig.\ref{photo}.

It was built as a test bench to develop the  optical and electronic technologies 
that will be implemented in GINGER. The latter is 
a very high sensitivity three dimensional array 
of large frame RLGs, currently in the design phase,
which has the aim of verifying general 
relativity theories on a laboratory scale \cite{angela2017}.  
In addition to this,  GP2 has also the aim of contributing to the development 
of  reduced scale RLGs (4 - 8 m in perimeter) with slightly lower performance, 
devoted to geophysical and seismic applications.

The quality of the performances of a RLG are strictly related to the control
of the geometrical scale factor $S=4 \frac{A}{P \lambda}\cos(\theta)$, 
and of the laser kinematics. 
In a previous papers we addressed the  complex   problem of the reduction 
of the non-linear laser dynamics perturbation \cite{Angela2019,Angela_epj2020}, 
elaborating a procedure that was applied to the $3.60\times 3.60~$m$^2$~  RLG
"Gingerino", placed in the deep underground Gran Sasso laboratory \cite{Belfi2017}.

In spite of its limitation, due to the reduced dimension and to the noisy location, 
GP2 can give important information about the procedures for the control 
of the scale factor, thanks to its peculiar properties. 
First of all, the ring  is oriented with its axis nearly parallel to the Earth axis,
so that the accuracy is improved and the disturbances induced on the Sagnac signal 
by local tilting are minimized. 
Furthermore,  the vacuum chamber includes also additional pipes connecting
the opposite mirrors along the two diagonals . 
In this way, the four mirrors define, besides the  square optical cavity, 
also two Fabry-P\'erot (FP) linear resonators along the  diagonals, 
that can be used to better define and control the ring geometry.

We already demonstrated in \cite{Stefani2018,Beverini_2020} 
to be able to measure and to stabilize the lengths of the diagonals of GP2
with a statistical accuracy of some tens of nanometers.
This results was obtained through an interferometric technique, 
in which a double Pound-Driver-Hall (PDH) control loop was realised for each diagonal arm.
In this work we investigate the geometrical properties of GP2, 
by testing new simplified procedures to measure and control the geometrical shape
and the scale factor that could be profitably applied to the reduced scale RLGs.

\section{The apparatus}

\subsection{The mechanical structure}

The optical cavity of GP2 is a square defined by four high-quality dielectric mirrors.
It is mounted inside a vacuum room composed by four corner chambers, hosting the mirrors,
connected, through bellows,   by pipes along the  sides and the diagonals of the square.
Each mirror is rigidly fixed to the chambers that are mounted on a piezoelectric 
slide (PZT), screwed on a massive granite table that ensures the stiffness
of the apparatus. 
 The PZTs can drive displacements along the  direction of the diagonals within 
 an  $80\ \mu$m range.
 
 The whole vacuum volume is filled by a 6.4 millibar of He and 0.2 millibar 
 of a mixture of $^{20}$Ne and $^{22}$Ne, with a 1:1 ratio of the two isotopes. 
 A pyrex capillary, 5~mm inner diameter, is inserted halfway in one side
where the gas is excited by a capacitive rf discharge.
In order to minimize the  losses and maximize the resonator quality factor,
no windows are inserted inside the laser optical paths.
A getter-pump minimises contamination of the noble gas mixtures.

The granite slab is fixed on a concrete base, tilted in the N-S direction 
by about $46.6^{\circ}$, which is the value of  the local colatitude. 
Besides, in order to have the ring axis oriented in the local meridian plane,
it was taken specific care of the positioning of the base .

\subsection{The optics}
~

We mounted very high quality mirrors 1" in diameter on GP2.  
Recently, we changed the optics with respect to that reported 
in Ref. \cite{Stefani2018,Beverini_2020}, mounting a more performing set of mirrors
and realigning the optical cavity.  
The new mirrors were tested, demonstrating at 45° s-polarized incidence 
a  reflectivity of  0.999995(1) and a transmission of 0.35(5) ppm,
in accordance with the supplier's specifications.
By means of ring-down measurements, the quality factor of the ring cavity
was obtained equal to approx$10^{12}$, corresponding to overall losses of 100~ppm. 
 Note that overall losses include also the losses for diffraction in the capillary.  
The reflectivity of the mirrors at normal incidence is somewhat lower,
so that the Q factor of the FP optical cavities along the diagonals is 
of the order of $10^{10}$. The high quality of the mirrors ensures
that even this small RLG is free from locking, as already demonstrated by \cite{stedman1993}
in a 0.75 m$^2$ RLG. 
Also the reference metrological laser (RfL) is presently
a Winters Electro-Optics Helium-Neon laser locked to a saturated absorption transition
of molecular iodine. Following the specification, the emission frequency is
$f_{Winter}=473 612 622.97\pm 0.01 $~MHz.
 By means of an optical hetherodyne technique between the RLG emission beam and 
 the reference laser on a fast photo-diode we can  measure the RLG frequency 
 in the new alignment with an accuracy of 1~MHz, limited by the RLG vibration noise.  

\subsection{Data acquisition and elaboration}

\begin{figure}[htbp]
\begin{center}
\includegraphics[width=\columnwidth]{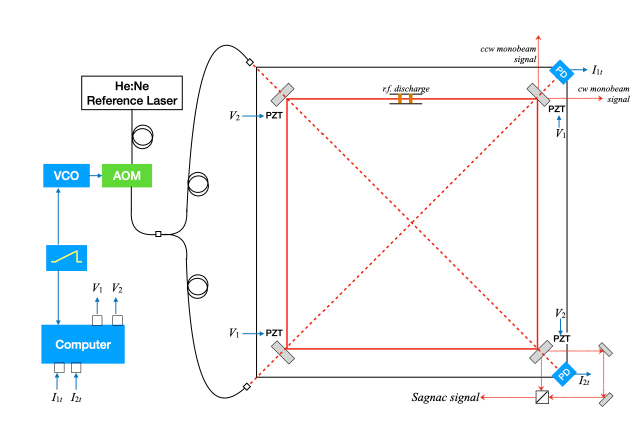}
\caption{Operational setup for geometrical stabilization.
          AOM Acoustic Optical Modulator;
           VCO Voltage Controlled Oscillator; PZT  piezoelectric device; PD photo diode; 
           $I_{1t}$, $I_{2t}$ transmission signal of the diagonal 1 and 2 resonators; 
           $V_1$, $V_2$ error signal of diagonal 1 and 2.}
\label{setup}
\end{center}
\end{figure}

The Sagnac interferometric signal is the beat note, obtained by a photo-diode, 
between  the clock-wise and counter-clock-wise beams.
The two beams are extracted by the same mirror, and the beat note is stored
continuously at a rate of 5 kSample/s.
In parallel at the same rate, we also record the intensity of 
the two counter-propagating beams ($monobeam~signals$) exiting from a second mirror, 
see Fig.\ref{setup}.
Data are  processed off-line by a MATLAB\textsuperscript{\textregistered}
procedure in order to correct 
the interferometric signal for mirrors back-scattering effects. 
Following the algorithms described in \cite{Angela2019}, 
we identify and correct these effects by using  the monobeam signals.
The procedure can be  executed quickly, producing the reduced data quasi on-line, 
in a few seconds, and includes the largest correction terms, 
usually referred as backscatter noise. 
In a second step, \cite{Angela_epj2020} linear regression methods could be applied 
to evaluate and correct the null-shift, related to the time evolution  
of the losses of the laser  optical resonator.
This procedure requires the off-line  processing of extended sets of data, 
of the order of  $10^4 - 10^5 $~s.
However, in RLG equipped with high quality mirrors the effects are quite small
with a slow evolution time, and can be neglected at the Fourier frequencies
of interest in seismic applications, frequency window spanning from 0.01 to 100~Hz.

\section{Stabilization of the RLG scale factor}

In order to stabilize the scale factor $S$, the standard techniques are to lock 
the ring  perimeter length by comparing the RLG emission frequency 
with an optical frequency standard or the RLG cavity FSR with a radio-frequency standard.
The presence of the two FP resonators along the square ring cavity allows 
the implementation of a procedure that optimise the ring shape and 
stabilize $S$ \cite{Santagata2015} by stabilizing the length
of the two diagonals.
Our procedure acts symmetrically on the PZTs, moving the four mirrors 
along the diagonal direction. It gives the advantage of keeping invariant the ring shape
and, as a consequence, the phase of the radiation back scattered by the mirrors, 
which is the principal source of error due the non-linearity of the laser kinematics. 

The new simplified scheme is shown in Fig.\ref{setup}.
The radiation coming from our RfL is sent through two polarization maintaining 
single-mode fibers into the Fabry-Perot resonators, 
 each formed by the presence of an opposing mirror.
Into the fiber path is inserted an acousto-optics modulator (AOM), 
operating around 200~MHz. 
Each FP transmission signal is collected on a photodiode,  then it's amplified
by a transimpendance modulus and finally it's sent to a data acquisition card 
National Instrument USB-6363.  
A VCO sweeps the in-fiber AOM every 100~ms of some MHz. 
A LabVIEW\textsuperscript{TM} program visualises the two transmission signals. At the beginning of each run, 
the PZTs that move the mirrors are manually driven by a DC voltage in order 
to center the transmission peaks inside the AOM sweep. Eventually, the sweep is reduced
to about 1~MHz and the LabView\textsuperscript{TM} program will calculate in two independent
parallel ways the voltage of the PZTs corresponding to the maxima 
of the transmission signals, sending the error signal to the PZTs to close the loop.
The high frequency response of the servo loop is limited to a few tenths of Hz,
because each PZT slide must move the whole chamber mass, bigger than 2~kg. 

We underline that the correction signal is sent symmetrically on the two PZTs 
of each diagonal. This solution  allows a correction of the temperature dilation
without  deforming  the geometry of the optical path, which would affect
the back-scattering phase.

\begin{figure}[htbp]
\begin{center}
\includegraphics[width=\columnwidth]{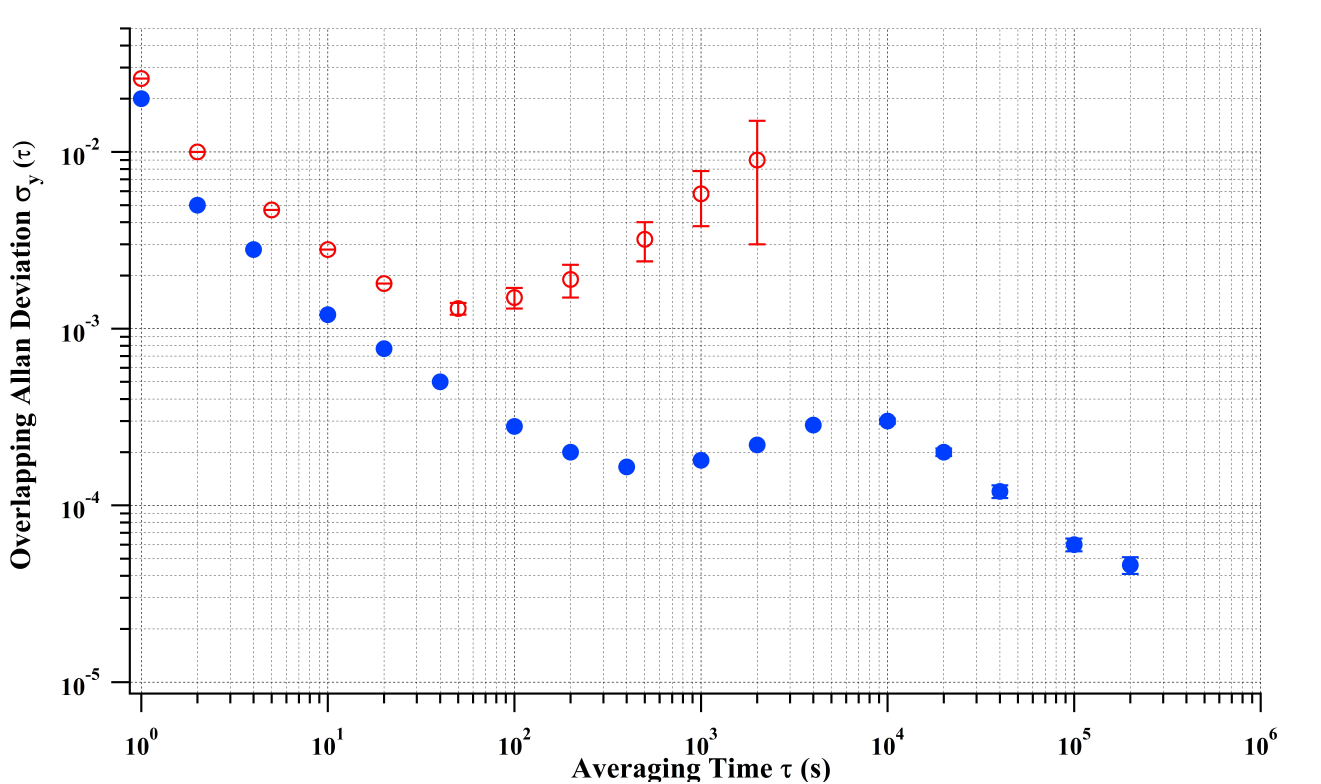}
\caption{
Allan plot of the Sagnac frequency    obtained from
a run with active locking condition 14 days long  (blue points) and 
to a selected free running  2-hours recording (red circles) 
          }
\label{Allan}
\end{center}
\end{figure}

\section {Discussion of the experimental results}

We compared Sagnac signal of two  2-days long runs,  recorded respectively with and
without the active stabilization of the diagonals, in slightly different environmental condition. 
In both cases,  the raw data were corrected to remove 
the backscattering contributions.
For the run without stabilization, we obtained a mean Sagnac frequency
value of 184.1593~Hz with a standard deviation of 0.0734~Hz, 
while for the locked diagonals run, we find
a mean Sagnac frequency value of 184.1327~Hz with a standard deviation of 0.0536~Hz.
Moreover, in locked condition no data point were lost
in the whole run, while in the free running case it was necessary to discard 
the 26\% of the data.

The efficacy of the stabilization is further demonstrated in Fig. \ref{Allan}  
that shows the Allan plot of the Sagnac frequency relative to a  14 days long locked run. 
No data were discarded and the Allan is certainly affected by the
environmental disturbances produced by the everyday activity.
At high frequency the signal is dominated by the environment vibration noise,
while it is achieved a long time stability better than 
$3 \times10^{-9}~$rad/s over one day.   
By comparison, we superimpose the Allan plot of a selected time interval 
(about two hours) in unlocked operation. Note that also in the best condition 
of the free running mode it was possible to acquire a continuous set of data
without spikes or jumps for no more than two or three hours. 

\begin{figure}
    \centering
    \includegraphics[width=\columnwidth]{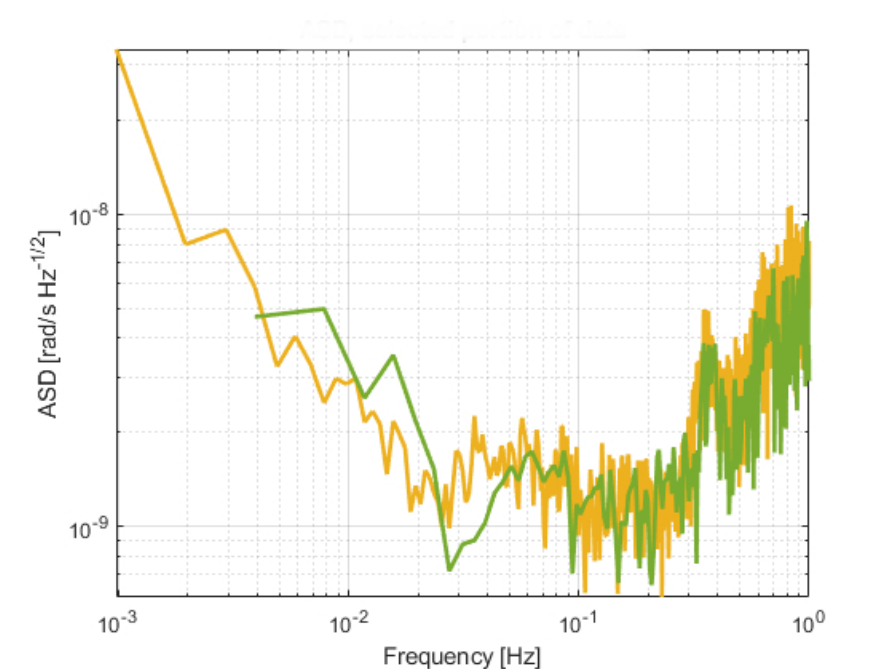}
    \caption{Amplitude Spectral Density calculated on two selected time intervals}
    \label{fig:ASD}
\end{figure}

The location guesting GP2 is a building downtown and it's quite noisy. 
There are different noise sources: electromagnetic one is related to the elevator,
vibrations to people moving through the corridor alongside our laboratory, sometimes not only people but also heavy instrumentation. 
We must stress that the apparatus is not simply affected by spurious rotations. 
Sagnac frequency is reconstructed from the interferometric signal by Hilbert algorithm and the accuracy of the procedure at the high frequencies is sensitive to the noise on the photodiode due to the electronic and to the vibrational noise. Equation 1 shows that the Sagnac frequency is affected also by the fluctuation of the orientation through the factor $\cos{\theta}$. 
We studied the effects of temperature variations in a high sensitivity RLG of 3.60 m 
of side placed in a very quit site, deeply underground 
in INFN GranSasso laboratory\cite{basti2021}.
An evaluation of its instrumental sensitivity in locked configuration can be deduced 
from the plots of Amplitude Spectral Density shown in  Fig. \ref{fig:ASD}, 
calculated in selected lime interval with the lowest environment noise.

A level of 2 nanorad/s/Hz$^{-1/2}$ is achieved in 0.02-0.3 Hz range. 
Note that in present configuration the high frequency locking efficiency
is limited by the PZT, as stated in section 3.

The technique described in the previous chapter allows the stabilization 
of the ring geometry and of the scale factor of the RLG, 
but it does not give the opportunity of optimizing the ring geometry following 
the procedure suggested in \cite{Santagata2015}. 
For this purpose it is necessary to measure the absolute values of the ring optical path
and of the diagonals length. Knowing this quantities, it is possible to evaluate
an expected value of the scale factor $S$ and to compare it with the measured one.  

The perimeter of the optical path can be calculated as $p=c / n\cdot $FSR, 
where  the refraction index of the intracavity medium $n$ can be estimated as
$1+2.1\times 10^{-7} $ \cite{Hurst2017}. 
The ring laser FSR is measured  by observing on a fast photo-diode 
the frequency spectrum of the RLG emission while increasing a bit 
the laser excitation up to the onset of a second longitudinal mode. 
We find a ring $p=6.3963216$~m.

To measure the absolute diagonal length we described in 
Ref \cite{Stefani2018,Beverini_2020} a complex procedure requiring for each diagonal
a double PDH locking circuit. Here we used an alternative more simple method 
by eliminating the second PDH locking. It is a 
less performing method, but it's accurate enough for our purpose.
As in \cite{Stefani2018}, the diagonal are measured one at a time in sequence.  
The light beam coming from the RfL is injected through polarization maintaining
single-mode fibers in the two diagonal resonators  and
 the single diagonal resonance frequency is locked to the RfL frequency 
 through a PDH circuit. 
 Then,  through an in-fiber EOM the laser radiation is frequency  swept 
 around a multiple of the resonator FSR. 
 A LabVIEW\textsuperscript{TM} software analyzes the transmitted signal fitting the data in order to find 
 the central FSR resonance frequency.
 The diagonal length is then given by $L=  c / (2n$~FSR).
 We found a value of the lengths of $2 261 341 \mu$m and $2 261 541~\mu$m, 
 with an estimated error of the order of 1~$\mu$m.
 
 We can observe that the perimeter $p'$ of a rhombus whose diagonals are equal 
 to the measured one, which is $p'=6.396~320$~m, is almost identical
 to the value $p=6.396~321~6$~m found before.

By using the observed values of the diagonals and of the perimeter,
we have also calculated the expected value of the scale factor $S$ of the GP2,
assuming a perfect rhomboid shape, perfectly aligned to Earth rotation axis, as:
$S=4A/(\lambda p)$
where $A$ is the area enclose by the optical path.
The intracavity wavelength $ \lambda $ is calculated as $c/f_0 $, 
where $f_0= 473 612 683$~MHz is the frequency emitted by the RLG, 
measured  beating on a fast photodiode the RLG radiation against the RfL.
Then, the scale factor of the RLG is $S=4A/(\lambda p)=2.52768\times 10^6$,
giving a theoretical Sagnac frequency due to the Earth rotation rate 
$\Omega_E = 7.29211586 \times 10^{-5}$ rad/s : 
$f_{th}=S  \Omega_E =184.216 Hz$. 

This can be compared with an experimentally detected Sagnac frequency of 184.133 Hz.
The difference could be consistent with a misalignment of the ring resonator axis
of $1.7^\circ $, but also with residual non compensation 
of the laser non-linear dynamics.

\section{Conclusions}

 Middle size RLG, such as GP2, are devoted to rotational seismology.
Rotational seismology requires instrumentation able to provide long term operation 
with sensitivity close to nrad/s and suitable response in the frequency range from 0.01 to 100~Hz. 
Our GP2 prototype, have a  simple and compact control apparatus and is
suitable to be installed in common geophysical observatories. 
This paper shows that they can satisfy the listed requirements.
The duty cycle of a RLG is limited in principle  by the dynamics of the laser, 
which is affected by mode jumps and split mode operation. 
The  duty cycle of our free running prototype is effectively less than $80\%$.
The control strategy, tested on GP2, allows to recover the $100\%$ of the duty cycle in 14 days long continuous unattended operation, without affecting the sensitivity response. 
Note that the present work make use of an expensive metrological laser with an accuracy better then $ 10^{-10}$ as reference length standard, but more simple commercially available reference lasers with a stability of 1 part on $10^7$  can provide a likewise  efficient control for geophysical application. The RLG has remarkable long term stability, even in the ultra low frequencies (corresponding to periodes up to a few hours), which are outside the range of classical geophysical instrumentation. Such stability can also open perspective for geophysics, with, for instance, possible applications in volcanology.

\section {Disclosure}
The authors declare no conflicts of interest.

\section{Data availability. }

Data underlying the results presented in this paper are not publicly available at this time but may be obtained from the authors upon reasonable request.

\bibliography{library}

\end{document}